\newcommand{\Msun}{~M_\odot}

\newcommand{\cmc}{\rm ~cm^{-3}}
\newcommand{\kms}{\rm ~km~s^{-1}}
\newcommand{\ergs}{\rm ~erg~s^{-1}}

\newcommand{\ml}{~\Msun ~\rm yr^{-1}}
\newcommand{\mll}{\Msun ~\rm yr^{-1}}
\newcommand{\Mdot}{\dot M}
\def\lsim{\raise0.3ex\hbox{$<$}\kern-0.75em{\lower0.65ex\hbox{$\sim$}}}
\def\gsim{\raise0.3ex\hbox{$>$}\kern-0.75em{\lower0.65ex\hbox{$\sim$}}}
\documentclass[12pt,preprint]{aastex}
\usepackage{epsfig}

\begin{document}

\title{THE DIVERSITY OF GAMMA-RAY BURST AFTERGLOWS AND 
THE SURROUNDINGS OF MASSIVE STARS}

\author{Roger A. Chevalier, Zhi-Yun Li}

\affil{Department of Astronomy, University of Virginia, P.O. Box 3818,  \\
Charlottesville, VA 22903; zl4h@virginia.edu, rac5x@virginia.edu}
\author{and Claes Fransson} 
\affil{Stockholm Observatory, Department of Astronomy, AlbaNova \\
S - 106 91 Stockholm, Sweden; claes@astro.su.se}

\begin{abstract}
The finding of a Type Ic supernova connected with GRB 030329
showed a massive star origin for this burst, supporting 
evidence for this association in previous bursts with lightcurve
bumps at the appropriate time for a supernova.
Here, we explore the possibility that all long bursts have
massive star progenitors, interacting with either the freely expanding
wind of the progenitor or the shocked wind.
We present models for the afterglows of GRB 020405 and GRB 021211,
which are a challenge to wind interaction models.
Considering sources for which wind interaction models are acceptable,
a range of wind densities is required, from values typical of
Galactic Wolf-Rayet stars to values $\sim 10^2$ times smaller.
The reason for the low densities is unclear, but may involve
low progenitor masses and/or low metallicities.
If mass is a factor, a low density event should be associated
with a low mass supernova.
The interpretation of bursts apparently interacting with
constant density media as interaction with a shocked wind
requires both a range of mass loss densities
and a range of external pressures.
The highest pressures, $p/k\ga 10^8$ cm$^{-3}$ K, may be due to
an extreme starburst environment, which would imply that the
burst is superposed on an active star forming region.
Although the range of observed events can be accomodated by the
shocked wind theory, special circumstances are necessary to
bring this about.
Finally, we consider the high velocity, high ionization absorption
features observed in some afterglow spectra.
If the features are circumstellar, the presence of the burst in a
starburst region may be important for the formation of clumps
near the burst.

\end{abstract}

\section{INTRODUCTION}

The spectroscopic finding of SN 2003dh in the afterglow light from GRB 030329
(Stanek et al. 2003; Hjorth et al. 2003) 
showed a direct link between a cosmological GRB
(gamma-ray burst) and the supernova explosion of a massive star,
supporting the previous identification of the nearby SN 1998bw with GRB 980425
(Galama et al. 1998).
This event also supports the view that the ``bumps'' found in
the light curves of a number of afterglows are, in fact, supernova light.
Sources with plausible supernova light from photometry include
GRB 970228 (Reichart 1999; Galama et al. 2000), GRB 980326 (Bloom et al. 1999),
GRB 011121 (Bloom et al. 2002; Garnavich et al. 2003), 
and GRB 020405 (Price et al. 2003b).
The recent GRB 021211 shows tentative evidence for supernova light
in HST (Hubble Space Telescope) observations
(Fruchter et al. 2002)
and evidence for a supernova spectrum like that of the normal Type Ic  SN 1994I
in the light of the burst has been reported 
(Della Valle et al. 2003).
In this case, the possible supernova is fainter than SN 1998bw and
is comparable in luminosity to  SN 1994I.

The implication of the supernova light is that the progenitor
object is a massive star.
The existing evidence is consistent with the supernova occurring
at the same time as the GRB, and that is the assumption that we make.
An implication of  a massive star progenitor is that the environment
for the progenitor is determined by the mass loss wind from the star
(M\'esz\'aros, Rees, \& Wijers 1998; Dai \& Lu 1998;
Chevalier \& Li 1999, 2000,
 hereafter CL99, CL00 respectively; Livio \& Waxman 2000).
If there is a free wind from the progenitor star, the wind
density is $\rho=Ar^{-2}=\dot M/4\pi r^2 v_w$, where $\dot M$
is the mass loss rate and $v_w$ is the wind velocity.
If the progenitor star is a Wolf-Rayet star, characteristic
mass loss parameters are $\dot M = 10^{-5}\ml$ and
$v_w=10^3 \kms$; the value of the density can be scaled to the corresponding
value for these parameters, $A_*=A/(5\times 10^{11} {\rm~ gm~cm^{-1}})$
(CL99).

The existing information on the afterglows of
GRB 970228 and GRB 980326 is consistent
with free wind interaction, although there is insufficient information
to tie down the wind density (CL99, CL00).
In the case of GRB 011121, wind interaction is indicated,
with $A_*\approx 0.02$  (Price et al. 2002c).
GRB 020405 presents more of a problem, because Berger et al. (2003)
found that a free wind model is not indicated and they proposed a
model involving interaction with a uniform medium.
For GRB 021211, Kumar \& Panaitescu (2003) consider both uniform
medium and wind models, both of which have certain difficulties.
For their wind model, the surrounding density must be low,
$A_*=5\times 10^{-4}$.
Because these two sources are of special interest from the massive
star progenitor point of view, we consider models for their
afterglows in \S~2.
In \S~3, we note that a number of sources appear to be better
modeled as interacting with a constant density medium, and
investigate the possibility that these objects involve interaction
with a shocked stellar wind (Wijers 2001).
In view of the evidence for a low density environment around some
GRBs, we consider the range of wind densities around Wolf-Rayet
stars in \S~4.
Another possible constraint on the environment of GRBs comes
from absorption lines observed in afterglow spectra.
Possible implications of the lines are discussed in \S~5.
Discussion and conclusions are in \S~6.
We do not consider the tantalizing evidence for X-ray lines,
which have not yet attained a high degree
of statistical significance.

\section{FREE WIND MODELS FOR AFTERGLOWS}

Although interaction with a free wind has been considered in
modeling afterglows, the number of objects for which such models
are successful remains small (e.g., Panaitescu \& Kumar 2001, 2002;
Yost et al. 2003).
Here, we consider two bursts that are of interest because they
have possibly been associated with supernovae.
The association is more secure in the case of GRB 030329, but
data are still becoming available for this object and the
optical light curve is complicated by fluctuations, so we
defer a discussion of this burst.

\subsection{Low-Density Wind Model for GRB 020405}
\label{grb020405}

The bright gamma-ray burst GRB 020405 occurred at April 5.0288, 2002 UT. 
It was detected and localized by the InterPlanetary Network (IPN) 
consisting of Ulysses, Mars Odyssey/HEND and BeppoSAX (Hurley et al. 
2002). The burst is a typical long GRB, with a duration of $\sim 60$~sec  
(Price et al. 2003b). An optical afterglow was identified by Price et al. 
(2002a) about 17 hours after the burst, which allowed for the redshift 
to be determined at $z=0.69$ later on (Masetti et al. 2002; Price et al. 
2002b). A late red bump was found in the optical data, which was 
interpreted as coming from an underlying supernova (Price et al. 2003b). 
Radio and X-ray afterflows were also observed (Berger et al. 2003; 
Mirabal, Paerels \& Halpern 2003a), making it possible to model the 
source.  

Berger et al. (2003) considered the multifrequency observations, and 
favored an afterglow model in which a jet expands into a constant 
density, presumably interstellar, medium. The evidence for the ambient 
medium being a constant density medium, as opposed to a massive 
star wind, comes from the radio data: the emission at 8.5~GHz is 
observed to have an initial flux of $\sim 0.5$~mJy at day 1.2  
and a rapid decline afterwards. The standard interpretation for 
such a ``radio flare'' is that it comes from the reverse shock of 
the GRB ejecta (Sari \& Piran 1999). This is possible only if the 
ejecta run into a medium of relatively low density, such as the 
typical interstellar medium. A dense Wolf-Rayet wind with a typical 
mass loss rate ($A_*\sim 1$)
can be ruled out (Berger et al. 2003), 
because the radio emission would die out on a time scale shorter than 
a day (CL00; 
Kobayashi \& Zhang 2003). It is, however, 
difficult to reconcile the constant density medium inferred in the 
immediate environment of this GRB with the aforementioned evidence 
for an underlying supernova: the latter points to a massive star 
progenitor, which should be surrounded by a wind. 

A possible resolution of the above paradox is that the progenitor wind 
has a  lower density (i.e., $A_*\ll 1$) than contemplated so far 
in the literature (see, however, Wu et al. 2003 and Dai \& Wu 2003). 
A low wind density would change the reverse shock evolution from
cooling to non-cooling and would prolong the radio emission from the reverse 
shock, making the wind model potentially compatible with the radio 
flare observed in GRB 020405. 

The optical and X-ray data provide some evidence for wind interaction. 
Bersier et al. (2003) found that the R band 
flux decreases with time approximately as $t^{-1.72}$ between 
day 1 and 4. The spectrum in the optical was determined to be 
$\nu^{-1.43\pm 0.08}$ at day 1.3, where a correction for Galactic extinction
has been included. 
The temporal slope in the optical, $\alpha_o
\approx -1.72$, agrees within uncertainties with that in the X-ray, 
$\alpha_X=-1.87\pm 0.1$, which is determined between about day 1.7 
and 2.3 from Chandra observations (Mirabal et al. 2003a). The 
spectral slope in the X-ray, $\beta_X=-0.72\pm 0.21$, is 
shallower than that in the optical, $\beta_o={-1.43\pm 
0.08}$ (Bersier et al. 2003) or $\beta_o\approx -1.25$ (Price et al.
2003b; see also Masetti et al. 2003), 
but agrees with the broad-band slope of $\beta_{oX}\approx 
-0.74$ between the optical and X-ray (Mirabal et al. 2003a). The 
agreement suggests that the steeper optical slope is probably caused 
by extinction in the host galaxy; some extinction is expected
for the host column density of $N_H=(4.7\pm 3.7)\times 10^{21}$
 cm$^{-2}$ determined from X-ray
observations (Mirabal et al. 2003a). 
If the optical spectrum has an intrinsic
slope similar to that of the X-ray spectrum, then the optical and X-ray data
can be explained naturally in a wind model with $p\approx 2.6$, 
provided that the cooling frequency $\nu_c$ is above the X-ray 
frequencies at the time of observation (Mirabal et al. 2003a). The
predicted temporal slope is $\alpha=-(3p-1)/4\approx -1.7$ and  
spectral slope $\beta=-(p-1)/2\approx -0.8$, both consistent with the 
optical through X-ray observations. The relatively high cooling 
frequency required in this model points to a low density in the 
progenitor wind and/or a low energy fraction for magnetic fields in 
the blast wave. 

To estimate the wind density, we will concentrate on modeling  
the emission from the forward shock, whose dynamic evolution is 
largely independent of the details of the initial energy injection 
from the central engine. The reverse shock, on the other hand, depends 
on the nature of the GRB ejecta, which is uncertain. The ejecta may 
have complex substructures and perhaps dynamically important 
large-scale magnetic fields (Coburn \& Boggs 2003), 
both of which complicate the dynamics of the 
reverse shock. 

To model the forward shock emission, we adopt the analytic expressions 
of Granot \& Sari (2002). These expressions are obtained by fitting 
numerical calculations based on the spherical, self-similar solutions 
of Blandford \& McKee (1976) for blast wave dynamics. The evolution 
of the nonthermal electrons injected at the shock front and their 
synchrotron emission are treated accurately. 

To begin the quantitative discussion, we first estimate the flux 
in the R-band ($\nu_R=4.5\times 10^{14}$~Hz) at $t=1$~day. Price 
et al. (2003b) determined a flux of $42.3\pm 6.2$~$\mu$Jy (after 
correcting for a Galactic extinction of $A_R=0.14$) at day 0.91. 
If the flux decays with time as $t^{-1.7}$, then the flux at
day 1 would be about 36~$\mu$Jy. We adopt a somewhat larger, round 
number of 40~$\mu$Jy, in view of the evidence for extinction in 
the host galaxy discussed above. This flux sets 
the overall scale for the emission in the optical and X-ray bands. 
It provides a constraint on the parameters of the wind model. From 
the expression for flux in Table~1 of Granot \& Sari (2002; segment G), 
we find  
\begin{equation}
A_*\ \epsilon_e^{8/5}\ \epsilon_B^{9/10}\ E_{52}^{9/10}
=5.8\times 10^{-5},
\label{e_scale}
\end{equation}
where a luminosity distance of $1.29\times 10^{28}$~cm (corresponding
to $z=0.69$ in a flat Universe with $\Omega_M=0.27$, $\Omega_\Lambda
=0.73$ and $H_0=71$~km~s$^{-1}$~Mpc$^{-1}$) and $p=2.6$ are adopted, 
and a subscript $n$ denotes a quantity divided by 
$10^n$. 
Here, $\epsilon_{B}$ is the fraction of the postshock energy density
in the magnetic field and $E$ is the equivalent isotropic energy
in the blast wave.
The reduced 
electron energy fraction ${\bar \epsilon}_e$ used in Granot \& Sari 
(2002) is here replaced by the more commonly used, actual fraction 
$\epsilon_e={\bar \epsilon}_e (p-1)/(p-2)$. 

A second model constraint comes from the radio emission observed after 
the initial flare. This component has a more or less constant flux 
of $\sim 10^2$~$\mu$Jy at 8.5~GHz up to about day~30 (Berger et al. 
2003). It could be produced in the forward shock since, in the slow
cooling regime under consideration, the flux at a given frequency 
is constant before the characteristic time $t_m$, when the typical 
frequency $\nu_m$ passes through that frequency (CL00). 
Denoting the constant flux produced in the forward shock by 
$F_{max,8.5{\rm GHz}}$, we find 
\begin{equation}
A_*\ \epsilon_B^{1/2}\ E_{52}^{1/2}=3.6\times 10^{-3} \left({F_{max,
8.5{\rm GHz}}\over 100\ \mu{\rm Jy}}\right)^{12/17}, 
\label{e_radio}
\end{equation}
where we have used the expressions for the typical frequency $\nu_m$ 
and the corresponding peak flux $F_{\nu_m}$ from Table~2 of Granot \& 
Sari (2002). Combining equations (\ref{e_scale}) and (\ref{e_radio}), 
we obtain an expression for the wind density parameter 
\begin{equation}
A_*=6.2\times 10^{-3} \left({\epsilon_e\over 0.1}\right)^2 
\left({F_{max,8.5{\rm GHz}}\over 100\ \mu{\rm Jy}}\right)^{27/17}. 
\label{e_density}
\end{equation}
Since the electron energy fraction $\epsilon_e$ is unlikely to exceed 
1/3 (the value for energy equipartition between nonthermal electrons, 
protons and magnetic fields) and $F_{max,8.5{\rm GHz}}$ must not be 
more than about $10^2\ \mu$Jy, we conclude that $A_* \la 6.9\times 
10^{-2}$. In other words, the density is well below than that of a 
typical Wolf-Rayet wind ($A_*\approx 1$). 

The low wind density is consistent with, but not required by, the high 
cooling frequency inferred from the optical/X-ray data. As mentioned 
earlier, the cooling frequency needs to be above the X-ray frequencies 
between day 1.7 and 2.3. Denoting the cooling frequency at day~1.7 
by $\nu_{c,1.7{\rm day}}$, we find from the expression for $\nu_c$ in 
Table~2 of Granot \& Sari (2002) that  
\begin{equation}
A_*^{-2}\ \epsilon_B^{-3/2}\ E_{52}^{1/2}=3.4\times 10^7 \left({
\nu_{c,1.7{\rm day}}\over 2.4\times 10^{18}\ {\rm Hz}}\right), 
\label{e_cool}
\end{equation}
where $\nu_{c,1.7{\rm day}}\ga 2.4\times 10^{18}$~Hz (corresponding
to an X-ray energy of 10~keV). This equation can be combined with 
equations~(\ref{e_scale}) and (\ref{e_radio}) to yield
\begin{equation}
\epsilon_B=2.1\times 10^{-2} \left({\epsilon_e\over 0.1}\right)^{-3}
\left({F_{max,8.5{\rm GHz}}\over 100\ \mu{\rm Jy}}\right)^{-69/34}
\left({\nu_{c,1.7{\rm day}}\over 2.4\times 10^{18}\ {\rm Hz}}
\right)^{-1/2}, 
\label{e_Bfrac}
\end{equation}
and
\begin{equation}
E=1.6\times 10^{53} \left({\epsilon_e\over 0.1}\right)^{-1}
\left({F_{max,8.5{\rm GHz}}\over 100\ \mu{\rm Jy}}\right)^{9/34}
\left({\nu_{c,1.7{\rm day}}\over 2.4\times 10^{18}\ {\rm Hz}}
\right)^{1/2}\ \ {\rm ergs}. 
\label{e_energy}
\end{equation}
The inferred spherical blast wave energy for the fiducial set 
of parameters ($\epsilon_e=0.1$, $F_{max,8.5{\rm GHz}}=100\ \mu{\rm 
Jy}$ and $\nu_{c,1.7{\rm day}}=2.4\times 10^{18}\ {\rm Hz}$) is
comparable to the $k$-corrected, isotropic-equivalent $\gamma$-ray 
energy release, which is $(7.37\pm 0.80)\times 10^{52}$~ergs according 
to Price et al. (2003b). 

In deriving equation~(\ref{e_cool}), we have assumed that inverse Compton 
scattering does not modify the cooling frequency significantly. We 
verified that this is the case for the fiducial parameters during the 
time of interest (after a day or so), when the forward shock is in 
the slow cooling regime and only a small fraction of nonthermal 
electrons radiate efficiently. If $\epsilon_e$ is much larger than
$0.1$, then $\epsilon_B$ would be very small, and inverse Compton scattering
could dominate synchrotron radiation in cooling the nonthermal electrons 
(e.g., Sari \& Esin 2001). In such a case, equation~(\ref{e_cool}) 
and expressions~(\ref{e_Bfrac}) and (\ref{e_energy}) need to be modified 
accordingly.

If the more or less 
constant radio emission at 8.5~GHz after the initial flare comes from 
the forward shock, then $F_{max,8.5{\rm GHz}}\approx 100\ \mu$Jy, 
which yields a self-absorption time 
\begin{equation}
t_a (\nu=8.5\ {\rm GHz}) \approx 0.02 \left({\epsilon_e\over 0.1}\right)
^{4/5}\left({\nu_{c,1.7{\rm day}}\over 2.4\times 10^{18}\ {\rm Hz}}
\right)^{-1/5}\ \ {\rm days}
\label{e_absorb}
\end{equation}
according to Granot \& Sari (2002). This time is before the first radio 
observation, and is consistent with it marking the beginning of the 
predicted constant-flux segment of radio light curves (CL00). 
The segment is expected to end at the characteristic time 
\begin{equation}
t_m (\nu=8.5\ {\rm GHz})\approx 97.4\ {\rm days},
\end{equation}
which is later than the observed turnover around day~30. We need to 
invoke an additional process, such as a jet break, to account for 
the turnover. A collimated blast wave is indicated by the high 
degree of polarization (up to $\sim 10\%$; Bersier et al. 2003) 
observed in the optical emission. To produce a break around day~30, 
the jet must have a half-opening angle of order (CL00) 
\begin{equation}
\theta_0\approx 0.06 \left({\epsilon_e\over 0.1}\right)^{3/4} 
\left({\nu_{c,1.7{\rm day}}\over 2.4\times 10^{18}\ {\rm Hz}}
\right)^{-1/8},
\end{equation}
which would bring the inferred blast wave energy down to 
\begin{equation}
E_{jet}\approx 3\times 10^{50} \left({\epsilon_e\over 0.1}\right)^{1/2} 
\left({\nu_{c,1.7{\rm day}}\over 2.4\times 10^{18}\ {\rm Hz}}
\right)^{1/4}\ \ {\rm ergs}. 
\end{equation}
This value is  in line with the ``standard'' energy advocated by 
Frail et al. (2001). 

Whether the observed radio flare itself can be explained by the free wind model
is uncertain. The flare presents two potential difficulties to the 
model. First, it may be absorbed in the forward 
shock, if the wind is dense enough. This is, however, not a problem 
since the forward shock becomes transparent to the 
radiation at 8.5~GHz well before the earliest time of observation 
according to equation~(\ref{e_absorb}). Second, after the reverse 
shock crosses the GRB ejecta, no new nonthermal electrons are accelerated, 
and the reverse shock emission cuts off above a certain frequency 
$\nu_{cut}$. The cutoff frequency is expected to be well below 8.5~GHz at 
day 1 or later for typical Wolf-Rayet wind parameters. In the case 
of GRB 020405, 
this may not be a problem for the following reason. As mentioned 
earlier, the optical through X-ray data can be explained if the cooling 
frequency $\nu_c$ at day~1.7 is comparable to, or greater than, $2.4
\times 10^{18}$~Hz. Since $\nu_c\propto t^{1/2}$, we have 
$\nu_c (t=60\ {\rm sec})=4.9\times 
10^{16}\ {\rm Hz} \times ({\nu_{c,1.7{\rm day}}/2.4\times 10^{18}\ 
{\rm Hz}})$ in 
the forward shock at the time $t_{cr}\approx 60\ {\rm sec}$ when the 
reverse shock crosses the GRB ejecta. Since the energy densities are 
comparable in the forward and reverse shocks at this time, the cooling 
frequency in the reverse shock (above which the emission cuts off) 
should be comparable to that in the forward shock if their magnetic 
energy fractions are also comparable. 
Kobayashi \& Zhang (2003) showed that the cutoff frequency 
scales with time as $\nu_{cut} \propto t^{-15/8}$ in a wind model
at $t > t_{cr}$, adopting simple assumptions about the GRB ejecta 
and their dynamics 
after the passage of the reverse shock front. The scaling yields a 
cutoff time $t_{cut}(\nu=8.5 {\rm GHz})=2.8\ ({\nu_{c,1.7{\rm day}}
/2.4\times 10^{18}\ {\rm Hz}})^{8/15}$~days, which is late enough 
to accommodate the radio flare observed at day 1.2. Between $t_{cr}$ 
and $t_{cut}$, the radio flux is predicted to decrease as $t^{-1/2}$, 
which is too shallow to explain the rapid decline of roughly $t^{-1.1}$ 
between the initial peak and the second observation at day~3.3. The 
cutoff time must therefore occur before day 3.3, which requires  
that the cooling at day~1.7 occurs close to 10~keV. The reverse shock 
emission at the lower frequency 1.4~GHz can extend beyond day~3.3. 
It may account for the unusual spectral slope of $\beta=-0.3\pm 
0.3$ observed between 1.4 and 8.5~GHz at day~3.3 when combined
with the radio emission from the forward shock, which is expected 
to have a spectrum $\nu^{1/3}$.

\subsection{Low-Density Wind Model for GRB 021211}
\label{grb021211}

GRB 021211 was an X-ray rich GRB detected and localized by HETE-2 on 
December 11.4712, 2002 (Crew et al. 2003). A bright optical afterglow 
was discovered and followed up within minutes of the burst (Wozniak 
et al. 2002; Li et al. 2003; Park et al. 2002; Fox et al. 2003). At an 
age of $\sim 90$~sec, the afterglow had an R-band magnitude of $\sim 1
4$ (Wozniak et al. 2002). It decayed with time approximately as $t^{-
1.6}$ for about 11~min. Thereafter, the decay flattened to roughly $t^{-1}$
(Li et al. 2003; Fox et al. 2003; Pandey et al. 2003). Vreeswijk et 
al. (2002)    
determined the redshift 
to be $z=1.00$. Late time observations show some evidence for a 
underlying supernova (Fruchter et al. 2002; Della Valle et al. 2003).

The close resemblance of the early light curve of GRB 021211 to that of 
GRB 990123 has led to the suggestion that the initial, faster declining 
emission comes from the reverse shock of the GRB ejecta (e.g., Li et al. 
2003; Fox et al. 2003), thought to be the case for GRB 990123 (Sari \& 
Piran 1999). The relatively shallow slope of this portion, $\alpha
\approx -1.6$, was taken as evidence against a wind model (Fox et al. 
2003). We note, however, that the earliest emission could be a 
combination of emissions from the forward and the reverse shock, and 
the shallow slope may or may not be a problem. This point will be 
discussed further toward the end of the subsection. Here we discuss 
the data after about 11~min in the context of a wind model, and try 
to constrain the wind density. 

The wind model predicts a decay slope of $\alpha=-(3p-1)/4$ below the 
cooling frequency $\nu_c$ and $\alpha=-(3p-2)/4$ above. In the R-band,
the decay observed after about 11~min is relatively flat, with $\alpha
\sim 1$. This slope would be difficult to accommondate in a wind model, 
unless the power index $p$ of electron energy distribution is close 
to 2 {\it and} the R-band lies above the cooling frequency. We choose 
an index slightly above 2, $p=2.1$, which yields a decay slope 
$\alpha=-(3p-2)/4=-1.075$ in
the cooling regime. The corresponding spectral slope is $\beta=-p/2=
-1.05$, which can be constrained by the color of the afterglow. 
The most relevant data come from the nearly simultaneous R and J 
band measurements of McLeod et al. (2002) and Bersier et al. (2002) 
around day 0.86. Their R-band magnitude of $23.20\pm 0.18$ and J-band 
magnitude of $21.73\pm 0.12$ yield a spectral slope of $\beta\approx 
-1$ using the photometric zeropoints of Campins, Rieke \& Lebofsky (1985) 
for J and Fukugita, Shimasaku \& Ichikawa (1995) for R, after correcting
for Galactic extinction of $E(B-V)=0.028$ (Pandey et al. 2003). This 
slope is consistent with the wind model within uncertainties. Based on 
B and K$_s$ band photometry, Fox et al. (2003) obtained a slope $\beta=
-0.98$ around day 0.1. This slope is again consistent with the model, 
although the K$_s$ band flux is measured at day 0.0823 while that at R 
band at day 0.1310; some uncertainties are involved in extrapolating the
data to a common time. Nearly simultaneous data exist at day 7 in J, H 
and K$_s$ band (Fox et al. 2003). They yield a spectral slope $\beta > 
0$, indicating contributions from other components than the afterglow
at this (late) time, possibly an underlying supernova. We conclude that 
while the sparse multiple-color data available are unable to provide 
strong support for the wind model, they do appear to be consistent with 
the model, provided that the R-band is emitted by fast cooling electrons 
up to about day~1 or beyond. 

As in the case of GRB 020405, we adopt the analytic expressions of 
Granot \& Sari (2002) to derive model constraints. The first 
constraint comes from the R-band flux, which has an estimated magnitude 
of 23.1 at day~1. From the expression for the flux in segment H of 
the lightcurves of Granot \& Sari, we find 
\begin{equation}
\epsilon_e^{11/10} \epsilon_B^{1/40} E_{52}^{41/40}   
=3.9\times 10^{-3}.
\label{Rflux}
\end{equation}
A second constraint comes from the characteristic time $t_{m,R}$ when
the typical frequency $\nu_m$ passes through the R-band. This should 
occur around, or before, the kink in the early light curve at $t\approx 
11\ {\rm min}=7.5\times 10^{-3}$~days; otherwise, there will be a flat 
segment in the light curve (with $F_\nu \propto t^{-1/4}$; CL00) 
between the early steep segment and the shallower segment 
at later times, which is not observed. This requirement translates 
to
\begin{equation}
\epsilon_e^2 \ \epsilon_B^{1/2}\ E_{52}^{1/2}=4.0\times 10^{-3} \left(
{t_{m,R}\over 7.5\times 10^{-3}\ {\rm days}}\right)^{3/2}, 
\label{e_earlybreak}
\end{equation}
where $t_{m,R}\la 7.5\times 10^{-3}\ {\rm days}$. From equation 
(\ref{Rflux}) and (\ref{e_earlybreak}), we can solve for $E_{52}$ 
and $\epsilon_B$ in terms of $\epsilon_e$:
\begin{equation}
E_{52}=5.1\times 10^{-3} \epsilon_e^{-1} \left({t_{m,R}\over 7.5\times 
10^{-3}\ {\rm days}}\right)^{-3/40},
\label{energy_e1}
\end{equation}
and
\begin{equation}
\epsilon_B=3.1 \times 10^{-3} \epsilon_e^{-3} \left({t_{m,R}\over 
7.5\times 10^{-3}\ {\rm days}}\right)^{123/40}.
\label{eB_e1}
\end{equation}

The afterglow of GRB 021211 was not followed up in X-ray. In radio,
it was searched for, but not detected (Fox et al. 2003). A stringent 
upper limit of $\sim 35\ \mu$Jy was obtained at 8.5~GHz by adding 
up data between day 8.9 and 25.8. This limit provides a further 
constraint on the wind model, 
\begin{equation}
A_*\epsilon_e^{-2/3}\epsilon_B^{1/3} E_{52}^{1/3}=6.0\times 10^{-3} 
\left({F_{max, 8.5{\rm GHz}}\over 35\ \mu{\rm Jy}}\right),
\label{Radioflux}
\end{equation}
where $F_{max, 8.5{\rm GHz}}\lsim 35\ \mu{\rm Jy}$. Combining equations
(\ref{e_earlybreak}) and (\ref{Radioflux}), we find 
\begin{equation}
A_*=0.24\ \epsilon_e^2 \left({F_{max, 8.5{\rm GHz}}\over 35\ 
\mu{\rm Jy}}\right) \left({t_{m,R} \over 7.5\times 10^{-3}\ 
{\rm days}}\right)^{-1}.
\label{den_e1}
\end{equation}

The fourth, and final, constraint comes from the time for the cooling 
frequency to cross the R-band. This cooling time is given by 
\begin{equation}
t_{c,R}=6.9\times 10^7 A_*^4 \epsilon_B^3 E_{52}^{-1}\ \ {\rm days}
=1.3\ \left({F_{max, 8.5{\rm GHz}}\over 35\ 
\mu{\rm Jy}}\right)^4 \left({t_{m,R} \over 7.5\times 10^{-3}\ 
{\rm days}}\right)^{5.3} \ \ {\rm days},
\label{coolcross}
\end{equation}
where equations (\ref{energy_e1}), (\ref{eB_e1}) and (\ref{den_e1})
are used to obtain the second equality. As discussed earlier, in 
order to explain the relatively slow decay of R-band lightcurve
in a wind model and the spectral slope observed at day 0.86, we 
need to demand that the cooling frequency cross the R-band later
than about day 1. This condition of $t_{c,R} \gsim 1$~day can 
be made consistent with the requirements that $F_{max, 8.5{\rm GHz}}
\lsim 35\ \mu{\rm Jy}$ and $t_{m,R} \lsim 7.5\times 10^{-3}$~days 
when and only when both $F_{max, 8.5{\rm GHz}}$ and $t_{m,R}$ are 
close to their upper limits, according to equation~(\ref{coolcross}). 
It implies that the cooling frequency crosses the R-band from below 
around day~1. After day~1, the light curve should steepen to 
$t^{-1.325}$ and the spectral slope should become flatter, with 
$\beta=-0.55$. These predictions are difficult to test, because 
afterglow data are sparse and scattered at late times, and  
may be contaminated by supernova light.

The inference that $t_{m,R} \sim 7.5\times 10^{-3}$~days has 
interesting consequences. It implies, from equation (\ref{eB_e1}), 
that $\epsilon_e$ must be greater than about 0.15; otherwise, the 
energy fraction in magnetic fields would be greater than unity. 
On the other hand, $\epsilon_e$ should be less than about 1/3,
as mentioned earlier. Picking a value between the two extremes,
$\epsilon_e=0.25$, we find that $\epsilon_B \approx 0.2$ 
according to equation~(\ref{eB_e1}), $E\approx 2\times 10^{50}
$~ergs according to equation~(\ref{energy_e1}), and $A_*\approx 
0.015$ according to equation~(\ref{den_e1}). The wind density 
could be somewhat higher, by a factor of two or so, for values
of $\epsilon_e$ near the upper limit. It is still well below that 
of a typical Wolf-Rayet star ($A_* \approx 1$). The inferred 
blast wave energy $E$ is relatively low, although not far from 
the ``standard'' value given in Frail et al. (2001). This energy 
is not corrected for jet effects, since there is no evidence for 
jet break in this source. The low energy is perhaps to be expected, 
given that GRB 021211 has one of the dimmest optical afterglows, 
and is not detected at radio wavelengths. 

Radio emission was searched for at 8.5~GHz around day 0.1, the 
earliest radio observation of any GRB to date. The source was 
not detected at this time, with a 
$3\sigma$ upper limit of 110~$\mu$Jy (Fox et al. 2003). The nondetection 
is consistent with the wind model for the following reason. As noted 
earlier, the 
reverse shock emission cuts off above a certain frequency $\nu_{cut}$ 
once all of the GRB ejecta have been shocked. We can estimate the cutoff 
frequency at day 0.1 as we did in the last subsection. Let the time 
when the ejecta are crossed by the reverse shock front be $t_{cr}=5\ 
t_5$~sec, where $t_5$ is of order unity for GRB 021211. From the 
inference $t_{c,R}\sim 1$~day and the scaling $\nu_c\propto t^{1/2}$, 
we obtain 
$
\nu_c(t=t_{cr})\approx 3.4\times 10^{12} t_5^{1/2} \ {\rm Hz}
$
for the forward shock. If the cooling frequencies in the reverse and
forward shocks are the same at $t=t_{cr}$, then from the scaling 
$\nu_{cut}\propto t^{-15/8}$ (Kobayashi \& Zhang 2003) we find that
$
\nu_{cut}(t=0.1\ {\rm days})\approx 2.9\times 10^6 t_5^{19/8}\ {\rm Hz}.
$ 
Therefore, at day 0.1, the observing frequency ($8.5\times 10^9$~Hz) 
is well above the cutoff frequency, by more than three orders of 
magnitude. 

A potential problem with the wind model is the relatively shallow decay
in the initial portion of the R-band light curve before $\sim 11$~min.
Fitting a single power-law through this portion yields $\sim t^{-1.6}$
(e.g., Fox et al. 2003; see also Pandey et al. 2003). This emission, if
coming from the reverse shock alone, would have too shallow a decay for
both the constant density and the wind model. The problem is more severe
for the wind model than for the constant density model. However, the
early afterglow is probably a combination of emissions from the forward 
and reverse shock, and one must subtract out the (uncertain) forward 
shock emission from the observed data to obtain the true reverse shock 
emission. One can in principle extrapolate the data at later times
backwards to accomplish this task. The fact that the flattening in the 
light curve around $\sim 11$~min is relatively mild, from $\sim t^{-1.6}$ 
to $\sim t^{-1}$ (Li et al. 2003; Fox et al. 2003; Pandey et al. 2003),
suggests that care must be taken in doing the subtraction. Because of 
the relatively modest contrast between the two fitted power-laws in the 
region of interest, the reverse shock emission is sensitive to how the 
subtraction is done, and can easily have a decay slope that is steeper 
than $t^{-2}$. For example, we have subtracted from the data before 
$t=7.5\times 10^{-3}$~days (or 11~min) a single power-law with a temporal 
index $\alpha=-1.075$ (as predicted in the wind model) that fits well 
the late time light curve and found that the residual light curve can 
be fitted by $t^{-2.2}$. 
Therefore, we believe that the relatively 
shallow initial decay in the R-band light curve may not be as serious 
a problem as initially feared, particularly in view of the uncertainties 
mentioned earlier in making predictions of reverse shock emission. 

A  lower value for the wind density, $A_*\approx 5\times 10^{-4}$,
was derived by Kumar \& Panaitescu (2003); they also found a higher
isotropic energy.
Our model and that of Kumar \& Panaitescu both
  use the optical flux at $t= 11$ minutes
and the evolution at that time to set a limit on $t_{m,R}$.
However, our model has $p=2.1$ (required for modeling the
optical after after 11~minutes) as compared to  $p=2.5$ for
Kumar \& Panaitescu, and they do not use the cooling constraint
that plays a role in our model.
In addition to the optical observations at $t\ga 11$ minutes, they
modeled the earlier emission as
reverse shock emission and the $\gamma$-ray emission itself
(assumed to be from the forward shock wave).
Both interpretations give a low circumstellar density, although
our result is less extreme than that of Kumar \& Panaitescu (2003).

\subsection{Results on Wind Density from Afterglow Modeling}

In Table 1, we list the afterglows that have been modeled assuming
interaction with a free wind ($\rho\propto r^{-2}$) density structure.
Panaitescu \& Kumar (2001, 2002; hereafter PK01, PK02,
respectively) have developed both uniform
medium and wind models for well-observed afterglows.
They find that only in the case of GRB 970508 is the wind model
preferred; in other cases (GRBs 991208, 991216, 000418), wind and
uniform medium models are equally acceptable.
Table 1 shows that most of the wind cases are compatible with
$A_*$ of $0.3-0.7$, which is in accord with wind densities for
Galactic Wolf-Rayet stars.

In the cases of GRBs 011121, 020405, and 021211, the wind density
is considerably lower.
Of the sources in the Table, these are also the best cases for the
presence of supernova emission.
There is a selection effect in that a lower density wind
leads to  weaker afterglow emission, in which supernova emission may be
more easily detected.
The case of GRB 021211 is especially striking in this respect; its
luminosity was lower than upper limits that have been set for a
number of other GRBs, indicating that bursts of this type may be
underrepresented in the sample.

An implication of these results is that Type II supernovae are unlikely
to be associated with a significant part of the afterglow sample.
The radio and X-ray emission from normal Type II supernovae show
that Type IIP events appear to have the lowest density circumstellar
wind and they have $\dot M\ga 10^{-6}\ml$ for $v_w=10\kms$
(Chevalier 2003), implying $A_*\ga 10$.
One event with a lower density immediately surrounding the supernova
is SN 1987A, but the low density extended out only a radial
distance $\la 3\times 10^{17}$ cm.
This inference is consistent with the finding that Type II supernova
progenitor stars are not suitable for the propagation of
GRB jets 
(MacFadyen, Woosley, \& Heger 2001; Matzner 2003)

\section{BURSTS IN UNIFORM DENSITY MEDIA}

Since early studies of well-observed GRB afterglows, interaction
with a uniform medium has generally been preferred over interaction with
a wind medium (PK02 and references therein).
The range of densities that is observed for these cases is typical
of that in the interstellar medium and one interpretation is that
these objects are interacting directly with the interstellar medium
of the host galaxies.
However, this is not expected for the environment of a massive star.
One possibility is that there is a separate kind of progenitor
object for these sources, such as a compact binary system.
Another possibility is that a constant density environment
can be created around a massive star.

The most plausible reason for a constant density around a massive
star is that the wind has passed through a termination shock,
creating a hot, approximately constant density region (Weaver et al. 1977).
Wijers (2001) has discussed this scenario for the interaction
regions of GRB afterglows (see also Ramirez-Ruiz et al. 2001);
the question is whether the observed afterglows can be accomodated
with conditions expected in a shocked wind.

\subsection{Pre-Burst Wind Interactions}

The shocked stellar wind depends on the evolutionary stages
prior to the Wolf-Rayet stage.
For Galactic stars, a standard evolutionary track is to start
as an O star, evolve through a RSG (red supergiant) phase 
or LBV (luminous blue variable) phase with
considerable mass loss, and ending as a Wolf-Rayet star (Garcia-Segura
et al. 1996a, 1996b).
At low metallicity, the RSG phase may be absent (Chieffi et al. 2003);
this may also be the case for certain binary stars.

We begin by considering the second case, where the star remains
blue throughout its evolution.
The total age of a massive star ($\ga 25\Msun$) is  
$t_*\sim 3\times 10^6$ years.
The typical wind velocity is $v_w=1000\kms$ and a typical
$\dot M=3\times 10^{-6}\ml$ leads to a reasonable amount
of mass loss during the evolution.
The wind power is then $L=0.5 \dot M v_w^2=9.5\times 10^{35}\ergs$,
so that $L_{36}=L/(10^{36}\ergs)$ provides a reference value for
the power.
If the surrounding pressure can be neglected, the radius of the
wind bubble is $R_b=0.88(Lt^3/\rho_0)^{1/5}=1.9\times 10^{20}
n_0^{-1/5}L_{36}^{1/5} $ cm at $t=t_*$ (Weaver et al.
1977), where $\rho_0$ is the ambient
density and $n_0=\rho_0/(1.67\times 10^{-24}{\rm~gm}){\rm~cm^{-3}}$.
The shock velocity is $V_b=0.6R_b/t=12 n_0^{-1/5}L_{36}^{1/5}\kms$.

The pressure associated with the shock front, $p=\rho_0 V_b^2$, 
can be expressed as $p_0=1.8\times 10^4 n_0^{3/5}L_{36}^{2/5}$, 
where we use $p_0$ for $p/k$ in units of K cm$^{-3}$.
If the surrounding pressure is larger than this, the expansion
of the bubble is slowed before $t=t_*$.
The interstellar pressure in the solar neighborhood, when various
components of pressure are included, is $p_0\sim 10^4$.
In Galactic molecular clouds, $p_0\sim 10^5$ (Blitz 1993).
High interstellar pressures are attained in an intense
starburst region, like the nuclear region of M82 where
$p_0\sim 10^7$ (Chevalier \& Clegg 1985).
In ultraluminous starburst galaxies, the pressure may reach
$p_0\ga 10^8$ (Chevalier \& Fransson 2001).
The evidence on the positions of GRBs in their host galaxies indicates
that they generally follow the light distribution, with some
bursts occurring in the very center of a galaxy and others
in a more peripheral position (Bloom, Kulkarni, \& Djorgovski 2002).
The bursts in the central, bright region of a galaxy are the ones
that are most likely in a starburst region and may be exposed to
a high pressure.

The density in the interstellar bubble depends on the uncertain
physics of heat conduction.
Heat conduction could be prevented by the magnetic field, which
is expected to be toroidal in the shocked wind region.
If it is, the density and pressure in the bubble are given by
(Weaver et al. 1977)
\begin{equation}
\rho\approx \left(16\over 15\right)^{3/2}{\dot M\over \pi R_t^2v_w},\qquad
p\approx \left(16\over 15\right)^{5/2}{3\dot Mv_w\over 16\pi R_t^2},
\label{bub}
\end{equation}
where $R_t$ is the radius of the wind termination shock.
The temperature in the bubble is $T\approx 10^7 v_8^2$ K.
Weaver et al. (1977) consider a particular case with heat
conduction ($L_{26}=1.27$, $v_8=2$, $n_0=1$, $t=10^6$ years) and find
that the temperature in most of the bubble is lowered to $10^6$ and
the density is correspondingly raised.
They suggest that the termination shock is not a sharp discontinuity.
However, it is possible that instabilities in the collisionless shock
suppress heat conduction.

In the case where the star goes through a RSG phase, the Wolf-Rayet
progenitor drives a shock wave into the slow moving wind; this forward
shock front is likely to be radiative.
The termination shock of the Wolf-Rayet star wind should be adiabatic,
leading to a relatively thick region of shocked wind.
The structure of the shocked region is discussed by Chevalier
\& Imamura (1983, hereafter CI83), who develop self-similar solutions 
and find that
in the strong shock limit (for both the RSG shock and the termination
shock), $R_t/R_c$ is determined by the mass loss rates of the winds,
$\dot M_{WR}/\dot M_{RSG}$, and the ratio of the wind velocities,
$v_{WR}/v_{RSG}$.
Using typical values for these parameters, $\dot M_{WR}/\dot M_{RSG}\approx
0.01-1$ and $v_{WR}/v_{RSG}\approx 100-200$, Fig. 5 of CI83
gives the ratios of the wind velocity to the
shock velocity of the outer and inner shock, their $b_1$ and $b_2$.
A typical value for $b_2$ is 10; i.e. the termination shock expands
at $v_w/10$ in the rest frame. From 
their Table 3, we  find that $R_t/R_c=0.4-0.7$, for the
above parameter ranges.
The density in the shocked region is approximately constant, although
it rises toward the contact discontinuity (Fig. 3a of CI83).
Beyond the contact discontinuity, a region of shocked RSG wind is expected.
Since it should have cooled to $10^4$ K or lower, a density jump
of $\ga 10^3$ is expected at this point because of approximate
continuity of the pressure.

The radius of the shocked region depends on the duration of the
Wolf-Rayet phase.  
As a reference value, we take $t_{WR}=3\times 10^4$ years.
For $b_2=10$, the radius of the termination shock is
$1\times 10^{19} v_8(t/t_{WR})$ cm and the pressure in the
shocked region is $p_0=2.5\times 10^5 \dot M_{-5}
v_8(t/t_{WR})^{-2}$.
In the case where the interstellar pressure is higher than this,
the expansion is stalled.

In order to investigate the case where the expansion is stalled by a
high interstellar pressure, we have carried out a one dimensional
numerical simulation using the VH-1 hydrodynamic code. The full
evolution of the wind bubble from the main sequence to the Wolf-Rayet
phase has been followed. As a representative case, we use a
mass loss history, wind velocity, and duration of the different
evolutionary phases typical of that of a 35 $\Msun$ star, as
calculated by Garcia-Segura, Langer, \& MacLow (1996b). The wind
velocity in the Wolf-Rayet phase is $1000 \kms$, and the mass loss
rate $1\times10^{-5} \ml$. 
For the pressure of the ISM, we take
$p_0=2\times 10^7$, which is  typical of a local starburst sample 
(Heckman, Armus, \& Miley 1990). The density of the ISM is taken to be $0.2
~{\rm cm^{-3}}$ and the temperature $\sim 5\times10^7$ K, but these
parameters are not important for the properties of the shocked bubble,
as will be explained below. Cooling is included, assuming an
equilibrium cooling curve. 

In Fig. 1, we show the structure of the wind bubble at the time of
the supernova explosion. The most interesting results of this
calculation are the dramatic decrease of radius of the termination
shock, $R_t\sim 0.4$ pc, and the increase in the extent of the
constant density bubble, $R_c/R_t\sim 4$, out to the dense red supergiant
shell. These numbers should be compared with $R_t\sim 14$ pc and
$R_c/R_t \sim 1.8$, respectively, in the case of a low pressure
ISM. Because the pressure in the shocked wind is  nearly in equilibrium
with the ISM, and the temperature $\sim 10^7 v_8^2$ K, the density in
the bubble is $\sim 0.5 (p_0/10^7)~ v_8^{-2} ~{\rm cm^{-3}}$,
independent of the mass loss rate and the ambient density. A large ISM
pressure will therefore result in a high value of the density in the
shocked bubble. The extent of the constant density region depends on
the duration of the WR stage, which varies with mass and metallicity,
and the ambient pressure. Considerations based on mass conservation
show that
\begin{eqnarray}
{R_c\over R_t} = && \left[1+2.57  
 \left({P v_w \over \Mdot_{\rm WR}}\right)^{1/2} t_{\rm
WR}\right]^{1/3} \cr \approx && 4.9~ \left({P/k \over 10^7 ~\rm K
 ~\cmc}\right)^{1/6} \left({v_w \over 10^3\kms}\right)^{1/6}
 \left({\Mdot_{\rm WR} \over 10^{-5}~\Msun}\right)^{-1/6} \left({t_{\rm
WR} \over 10^5 ~\rm years}\right)^{1/3}.
\label{eqrsrt}
\end{eqnarray}
For a $40\Msun$ star without rotation, the Wolf-Rayet lifetime is
found to be $\sim 5.4\times10^5$ yrs at $Z = Z_\sun$, while it is only
$\sim 0.81\times10^5$ yrs at $Z = 0.4 Z_\sun$ (Maeder \& Meynet
1994). The structure of the dense red supergiant shell at 1.7 pc
should only be taken as a rough approximation. As the calculations by
Garcia-Segura et al. (1996b) 
show, this region is severely distorted by hydrodynamical
instabilities.

\subsection{Observed Bursts}

Several GRB afterglows have been inferred to be surrounded by
constant density regions.  The problem is to determine whether
these regions are consistent with expectations for a shocked
stellar wind.
Table 2 lists 4 bursts that have measured redshifts and have been
identified as likely to be interacting with a constant density medium.
The estimates of the density  are from
PK02.
Harrison et al. (2001) have modeled GRB 000926 and obtained
$n\sim 30$ cm$^{-3}$, in good agreement with PK02.
Piro et al. (2001) obtained $n\sim 4\times 10^4$ cm$^{-3}$ for this
source, but their model does not include the radio data.
Other sources modeled by PK02 can also be fit by a constant density
model, but an $r^{-2}$ wind model can also fit the data.
In any case, the densities inferred for GRB 990123 and GRB 000926
span the range for the sources.

The density and pressure in an adiabatic wind bubble (eq. [\ref{bub}])
can be used to obtain the relation $p=\rho v_w^2/5$ in the bubble.
The pressure in the bubble can thus be written as
$p_0=2.4\times 10^7 nv_8^2$, leading to the values of
$p_0 v_8^{-2}$ given in column (6) of Table 2.
The relatively low value of $p_0$ for GRB 990123 is appropriate
for a wind bubble in an interstellar medium with a moderate pressure.
This is consistent with the position of GRB 990123 away from the
the brightest regions of star formation (Bloom et al. 2002).
However, the high pressures inferred for GRB 000301c and GRB 000926
are difficult to reconcile with a plausible pressure expected in
an adiabatic wind bubble unless the bursts are occurring in an
extreme starburst region.
GRB 000301c and GRB 000926 are centrally located in their host galaxies,
although not perhaps at the very center (Bloom et al. 2002).
Heat conduction gives the possibility of increasing the density
and decreasing the temperature of the gas (Weaver et al. 1977).
However, we believe that the toroidal magnetic field in the wind
is likely to inhibit conduction.
In addition, the radiative cooling time for the gas becomes less
than the likely evolution time.
The cooling time for gas with $n\sim 30$ cm$^{-3}$ and $T\sim 10^7$ K
is $10^5$ yr; at $T\sim 10^6$ K, the cooling time is decreased by
the lower temperature and the enhanced radiation by CNO ions.

X-ray observations of bubbles around Galactic Wolf-Rayet stars have
the potential to shed light on the question of bubble properties.
Two bubbles, NGC 6888 (Wrigge et al. 1998) and S308 (Chu et al. 2003),
have been detected at present.
NGC 6888 shows a hot component ($8\times 10^6$ K) which requires
reduced heat conduction, but the component is more limb brightened
than expected in standard models (Wigge et al. 1998).
The observations indicate some complexities in the bubble
structure, but they do not clearly show the nature of the medium
occupying most of the volume.

In our scenario, any source with an inferred surrounding density
$\sim 30$ cm$^{-3}$ is expected to be in an extreme starburst region.
Frail et al. (2003) recently modeled the afterglow of GRB 980703 as
interaction with a constant density medium with $n\sim 30$ cm$^{-3}$;
a wind model is less likely, although it cannot be ruled out.
Berger, Kulkarni, \& Frail (2001) found radio emission from
the host galaxy of GRB 980703 that implies it is in the class of
ultraluminous infrared galaxies, although at the faint end.
The burst is near the center of the galaxy in a region of star formation,
so a high starburst pressure is possible.

Another constraint on possible  surrounding wind structure comes
from the fact that the afterglow shock front must reach the 
termination shock by the time, $t_i$, that the initial observations
used in the model have been made.
Column (5) of Table 2 gives estimates of this time for the PK02 models,
corrected to the source frame.
For an adiabatic wind bubble, the radius of the termination
shock can be estimated from eq. (\ref{bub}), yielding 
$R_t=1.1\times 10^{18} (\dot M_{-5}/v_8)^{1/2}n^{-1/2}$ cm;
the resulting values of $R_t$ for the different densities are
given in column (7).
An estimate of the time that it takes the shock front to reach
$R_t$ depends on energy per unit solid angle of the initial 
jet propagation before deceleration and spreading.
For spherical expansion, the shock radius is
\begin{equation}
R=4.9\times 10^{17} E_{53}^{1/2}(\dot M_{-5}/v_8)^{-1/2}t_d^{1/2} {\rm cm}, 
\label{rad}
\end{equation}
where
$E_{53}$ is the isotropic energy in units $10^{53}$ ergs
and $t_d$ is the age in days (CL00).
Column (4) of Table 2 gives an estimate of the initial isotropic
energy for the energy and jet angle given by PK02.
The requirement that the shock front reach $R_t$ before $t_i/(1+z)$
gives an upper limit on $\dot M/v_w$, which is in column (8) of Table 2.
It can be seen that the sources with a high estimated pressure have
values of $\dot M_{-5}/v_8$ in the expected range, but that the value for
GRB 990123 is  low.
For a standard Wolf-Rayet star wind velocity, the mass loss rate
must be  low in this case in order to accomodate a value of $R_t$ so
close to the star.

A final constraint comes from the duration of the afterglow that
can be described by spherical expansion in a constant density
medium;
during this time, the relativistic shock expands as $R\propto t^{0.25}$.
The maximum initial time of this expansion is $t_i$ and the
final time is $t_f$, which for these sources occurs when jet
deceleration and lateral expansion become important.
This time    
is given by Frail et al. (2001) for the
sources considered here.
The maximum value of the radial range $R_i/R_f$ is thus
about $(t_i/t_f)^{0.25}$ and is given in column (9) of Table 2.
The maximum values are in some cases close to the minimum values
that are expected if the Wolf-Rayet wind is expanding into
a RSG wind and the external pressure is low, 
so that some interaction models can be ruled out.
The case of a high external pressure can give a more extended shocked
bubble, which provides a more suitable region for the constant
density expansion.

The above considerations show that there are possible problems with a shocked
Wolf-Rayet wind interpretation of the results from GRB afterglow models.
The low density and early observations of GRB 990123 require a
surprisingly low value for the progenitor mass loss rate,
although it is similar to that deduced for GRB 021211.
The high densities inferred for GRB 000301c, GRB 000926, and
some other bursts imply
 pressures that are only expected in an extreme starburst region.
In addition, the time range of observations suggests a radial range
for the shock wave of a factor $>2$ in radius.
Interaction with either the termination shock density
jump (factor 4 in density) or the contact discontinuity density
jump (factor $\ga 10^3$ in density) might be expected in some cases.
Wijers (2001) suggested that the bump in optical afterglow of
GRB 970508 was due to interaction with the wind termination shock.
However, the afterglow evolution before the jump should be steeper
or the same as the evolution after the jump; that is not observed
for GRB 970508 (e.g., Fruchter et al. 2000).
Another explanation for the bump is emission from a collimated
flow that is initially directed away from the observer (PK02).
In the case of GRB 030226, Dai \& Wu (2003) suggested that the
apparent increase in flux along with a steepening of the light
curve is due to interaction with the contact discontinuity, but
the sparse data do not make the case completely clear.

In addition to placing constraints on models with constant density
interaction, the shocked wind scenario constrains the free wind
models in that the termination shock must lie beyond the region
where the free wind model has been applied.
Using eqs. (\ref{bub}) and (\ref{rad}), this constraint can be
expressed as an upper limit on the pressure surrounding the wind:
\begin{equation}
p_0<1.3\times 10^8 A_*^2 v_8^2 E_{53}^{-1} t_f^{-1},
\end{equation}
where $t_f$ is the time of latest application  of the spherical afterglow
model in days.
In the case of our model for GRB 021211 ($A_*=0.015$, $E_{53}=0.002$,
$t_f=30$), we have $p_0<5\times 10^5 v_8^2$.
The model requires that the burst did not   occur in an intense
starburst region.
The wind model for GRB 021211 discussed by Kumar \& Panaitescu (2003)
($A_*=5\times 10^{-4}$, $E_{53}=3$,
$t_f=30$) yields $p_0<0.4 v_8^2$, an implausibly small value for the
external pressure.
Kumar \& Panaitescu (2003) argue against the wind model based on the
low value of $A_*$ and the high value of $\epsilon_B$; the inability
of the wind to propagate out from the star 
in their model provides another reason.
Kumar \& Panaitescu (2003) prefer a constant density model with
$n\sim 10^{-2}-10^{-3}$ cm$^{-3}$; as in the case of GRB 990123,
this model will require a low wind density and surrounding
pressure, if the constant density region is a shocked wind.

\section{LOW DENSITY WINDS AROUND WOLF-RAYET STARS}

The observations of GRB afterglows indicate that, in some cases,
the densities surrounding the bursts are lower than expected around
  Wolf-Rayet stars, which are believed to be likely progenitors
of the bursts.
The observational support for Wolf-Rayet progenitors has come
from the identification of GRB 980425, GRB 030329,
and possibly GRB 021211 with Type Ic
supernovae (Galama et al. 1998; Stanek et al. 2003; 
Hjorth et al. 2003; Della Valle
et al. 2003).
The main theoretical argument in their favor is the need to have
a compact stellar progenitor in order to get the explosion out
from the central engine in an amount of time  that is not much
longer than the duration of gamma-ray bursts (MacFadyen, Woosley, 
\& Heger 2001; Matzner 2003).

Mass loss rate estimates for Galactic Wolf-Rayet stars have been 
given by Nugis \& Lamers (2000), who find typical mass loss rates
$\dot M \approx 10^{-5.4}-10^{-4}\ml$ and wind velocities
$v_w = 700-5000\kms$.
The mass loss rates given by Nugis \& Lamers include a correction
for clumping, which brings the rates down by a factor of a few compared
to the uncorrected values.
Of the stars in their list, the one with the lowest wind density 
is a WO 2 star with
$\dot M = 0.39 \times 10^{-5}\ml$ and $v_w = 5500\kms$, yielding $A_*=0.07$.
The high wind velocity of a compact WO star is a significant part 
of the low density.

Possible reasons for a significantly lower mass loss density are a
metallicity dependence, mass dependence, and azimuthal dependence
of the mass loss.
Wijers (2001) argued that GRBs occur in low metallicity galaxies
so the stars have lower mass loss rates, with $\dot M \approx 10^{-6}\ml$.
However, WN type Wolf-Rayet stars have been found in the lower metallicity
Large Magellanic Cloud (Hamann \& Koesterke 2000)
 and the Small Magellanic Cloud (Crowther 2000)
that have mass loss rates that are comparable to Galactic stars.
However, in a similar study of WC stars, Crowther et al. (2002)
found that the mass loss rates scale as $\dot M\propto Z^{0.5}$,
where $Z$ is the metal fraction, similar to that found for
main sequence stellar winds.
The mass loss mechanism for Wolf-Rayet stars is poorly understood.
WC stars presumably have comparable amounts of element processing
despite different initial conditions.
If the mass loss depends on these elements, the dependence on
initial metallicity may be small.
However, if Fe lines are important, the initial metallicity
plays a role.

On evolutionary grounds, Langer (1989) advocated 
$\dot M =(0.6-1.0)\times 10^{-7}  
(M_{\rm WR}/\Msun)^{2.5}$  $\mll$, where $M_{\rm WR}$ is
the mass of the Wolf-Rayet star.
The stellar mass drops to $5-10 \Msun$ at the end of its life because of
mass loss, so $\dot M\sim (0.4-3)\times 10^{-5}\ml$ at that time.
With this prescription for mass loss, Wolf-Rayet stars are expected
to end up in a fairly small final mass range.
The mass loss rates of Nubis \& Lamers (2000) are a factor $\sim 2$
lower than those used by Langer, which leads to a larger final
mass.  
However, it is not clear whether the present evidence supports a
small final mass range for Wolf-Rayet stars.
The mass of ejecta in SN 1998bw is estimated to be
$10\Msun$, with a presupernova mass of  
$13.8\Msun$ (Nomoto et al. 2001).
On the other hand, the mass of ejecta in SN 1994I has been
estimated at $0.9\Msun$, with a presupernova mass of $2.1\Msun$
(Nomoto et al. 2001).
Although there is   uncertainty in these estimates, the case for
SN 1994I having a considerably lower mass than SN 1998bw is good.

If low progenitor mass is the reason for a low density wind, there should be
a relation between the supernova characteristics and the presence
of a low density wind, although variations in the explosion energy
are another factor in the diversity of supernova properties.
A low progenitor mass leads to a fast (rapidly evolving) supernova,
as was observed for SN 1994I (Nomoto et al. 2001).
The estimated energy for the explosion of SN 1994I is 
$1\times 10^{51}$ ergs.
If the energy were high, $\ga 10^{52}$ ergs as indicated for SN 1998bw
(Nomoto et al. 2001), the light curve would be especially fast
because the timescale for the lightcurve 
$\sim E^{-1/4}\kappa^{1/2}M_{ej}^{3/4}$.
In the case of GRB 021211, the evidence for a supernova shows that
it is comparable in luminosity and spectrum
to SN 1994I and is not bright like
SN 1998bw (Della Valle et al. 2003).
Unless there is a large compact remnant mass, 
this implies a low mass progenitor, which is consistent with
the low value of $A_*$ deduced for GRB 021211.  
In the case of SN 2001ke associated with GRB 011121,
the supernova is fainter and faster than SN 1998bw
(Bloom et al. 2002; Garnavich et al. 2003), intermediate
between SN 1998bw and SN 1994I.
This argues for an intermediate progenitor mass.

In standard models for GRBs in supernovae, the burst occurs along
the stellar rotation axis (e.g., MacFadyen, Woosley, \& 
Heger 2001), so a lower wind density
along the polar axis compared to the equatorial region would be
observed as a low density interaction.
This kind of structure was proposed in a model in which rotation and
magnetic fields play a crucial role in the mass loss from
Wolf-Rayet stars (Poe, Friend, \& Cassinelli 1989;
Ignace, Cassinelli, \& Bjorkman 1998).
In this picture, the radiation pressure-driven mass loss in the polar
direction would be comparable to that observed in O stars and would be
considerably less than the equatorial mass loss.
However, a prediction of this model is polarization of the light
in scattering lines, which is not observed in most Wolf-Rayet stars
(Harries, Hillier, \& Howarth 1998; Kurosawa et al. 1999);
in those stars that do show polarization, the required equator:pole
density ratio is $\sim 2-3$ (Harries et al. 1998).
Aspherical winds appear to be unlikely to explain the low values
of wind density required for some GRB afterglows. 

GRBs may represent only a small fraction ($\la 10^{-2}$) of Wolf-Rayet
star deaths, so there is the possibility that they have unusual
conditions that are not seen in observed stars.
A factor in producing a GRB is thought to be stellar rotation
in order to have a rapidly rotating core, so rotation may be
a characteristic of the progenitors.
The models of Meynet \& Maeder (2003) show that the rotational velocities
for the Wolf-Rayet stars are likely to be low, $\sim 50\kms$, but
stars in binary systems may be faster rotators.
However, rapid rotation may affect the mass loss in the equatorial
plane, but mass loss in the polar direction, which is the direction
thought to be relevant to GRB flows, may not be affected.

\section{OPTICAL/ULTRAVIOLET ABSORPTION LINES}

One of the interesting recent observations of GRB afterglows
is the finding that some of them show absorption lines of highly ionized
species, blueshifted relative to the host galaxy.
The best studied of these is GRB 021004 (Mirabal et al. 2002, 2003b;
Salamanca et al. 2002; Saviglio et al. 2002; Schaefer et al. 2003).
Mirabal et al. (2003b) find a host redshift of $z=2.328$ and
absorption lines of CIV, SiIV, and Lyman lines at 
$z=2.323, 2.317, 
2.293 $, corresponding to velocities of $-450, -990, -3155\kms$
relative to the host.

Schaefer et al. (2003) and Mirabal et al. (2003b) have discussed
the possible origin of the blueshifted features and have both
concluded that a circumstellar origin is most likely.
There are two possibilities for the velocity structure:
the high velocity is representative of the Wolf-Rayet star
wind velocity and the lower velocities are representative
of shells swept up by the Wolf-Rayet wind, or the absorption
features are due to nearby clumps that are radiatively
accelerated by the radiation from the GRB.
Intermediate situations between these two are also possible.
Schaefer et al. (2003) prefer the first interpretation because
velocities in the general range of those observed are
naturally produced.
Mirabal et al. (2003b) argue against this interpretation and
for radiative acceleration, based on the abundances in the
fast component.
H is definitely present; although Wolf-Rayet stars are H poor,
some WN stars have H, as do a significant
fraction of Type Ib supernovae (Branch et al.
2002).
The abundances deduced by Mirabal et al. do not show the
 overabundance of N expected for a WN, so they conclude against
 this possibility.
However, the abundances are very uncertain, given that only high
ionization stages are observed, so this argument may not be conclusive.

In addition to GRB 021004, high excitation, high velocity absorption
features have been found in GRB 020813 (Barth et al. 2003) and
GRB 030226 (Greiner et al. 2003; Price et al. 2003a; Chornock \& Filippenko 2003).
The absorption lines of CIV in GRB 020813 are at $z=1.223$ and
$z=1.255$ (Barth et al. 2003); the $z=1.255$ system is also present
in a [O II] emission line, indicating that this is the redshift of
the host galaxy.
The $z=1.223$ system has a velocity of $-4320\kms$ relative to the host.
In this case, the blueshifted absorption is also present in a number
of lower ionization species (Si II, Al II, Fe II, Mg II, and Mg I);
there is no coverage of Ly$\alpha$.
In the case of GRB 030226, strong absorption line systems are present
at $z=1.961$ and $z=1.984$, with C IV and Si IV present, as well
as numerous lower ionization species and Ly$\alpha$.
The velocity separation is $2300\kms$.
The velocity separation seen in these sources is consistent with
expectations for the velocity of a Wolf-Rayet star wind.
However, the presence of H does not support this origin for
the lines.

The main problem with a circumstellar origin for the absorption
is the strong ionization by the GRB radiation.
Lazzati et al. (2002) estimate that C IV is photoionized around
GRB 021004 out to $\ga 10^{19}$ cm,  which is larger than the
radii at which radiative acceleration would have to occur ($\sim 10^{18}$ cm).
A possible way to regain the C IV is recombination, but the 
density required to recombine to C IV is $\sim 10^7$ cm$^{-3}$
(Lazzati et al. 2002), which is much larger than the density
expected in a Wolf-Rayet star wind ($<1$ cm$^{-3}$).
The required degree of clumping in the Wolf-Rayet star wind itself
does not seem plausible, but there is the possibility of clumps from
the swept up RSG wind.
The 2-dimensional simulations of Garcia-Segura et al. (1996b) show that
the swept up RSG wind is fragmented and spread over a radial range.
The material ends up at a radius of  $>4$ pc in the simulations of
Garcia-Segura et al., but the high pressure in a starburst region
can keep the RSG wind shell at a relatively small radius.
This scenario also has the benefit of providing a high pressure to compress
the RSG clumps.
With the case $p_0=10^8$, clumps that may be shielded from the Wolf-Rayet 
star radiation field may have a high enough density for recombination
to be important.

Photoionization may also be a problem for line formation in the
circumstellar wind, but if the wind is relatively strong, the
free wind may extend to $\ga 10^{19}$ cm from the star.
If the free wind is the source of the high velocity absorption,
there are implications for the afterglow observed from the GRB.
The afterglow evolution must be of the wind type over a period
of at least days and the wind
must be strong enough to sustain a larger termination shock radius.
These expectations are borne out for GRB 021004, for which
Li \& Chevalier (2003) deduced a free wind type model with
$A_*=0.6$.

\section{DISCUSSION AND CONCLUSIONS}

In this paper, we have emphasized some consequences of the
hypothesis that all the long duration GRBs have massive star
progenitors.
Because the massive stars are expected to have their close-in
surroundings modified by the progenitor winds, we consider
both free winds and shocked winds as possible surrounding media
for the afterglow phase.
The properties of a shocked wind can be affected by the pressure
of the surrounding medium when the pressure is high, as occurs
in a starburst region.
A number of afterglows have been interpreted in terms of interaction
with a constant density medium with density $\sim 20-30$ cm$^{-3}$
(Yost et al. 2003; Frail et al. 2003).
Yost et al. note that this density is compatible with that of
interstellar clouds or the interclump medium of Galactic molecular clouds.
A different interpretation is needed in the massive star progenitor
case, and we investigated the possibility that the medium is the
shocked wind bubble in a starburst region.
The relatively high densities found in some afterglows require that
the burst occur in an extreme starburst region.
In the case where there is direct interaction of a GRB with the
interstellar medium, the interclump medium in a molecular cloud
in a starburst region
is likely to have a considerably higher density than the densities
observed ($\ga 10^3$ cm$^{-3}$, Chevalier \& Fransson 2001).
A significant fraction of GRBs must occur in starbursts,
so the fact that such high densities have not been observed
may be an indication that the surroundings of GRBs have been
modified by stellar winds or ionizing radiation.

The fact that there is a substantial number of bursts that require
a very high pressure surrounding medium may be a drawback to this
scenario, but such sources are likely to be overrepresented in
the sample because they are luminous and can be observed in detail.
Among the afterglows that can be interpreted as interaction with 
a free wind, the highest density objects are compatible with 
expectations for the wind from a typical Galactic Wolf-Rayet star,
but the lower densities imply a wind densities that are lower by
a factor $\sim 100$.
Because the density in a shocked wind is higher than that in a
free wind at the same radial point, the low density requirement
is not alleviated by appealing to a shocked wind.
One factor contributing to a low density wind may be a high
wind velocity, as appears to occur for WO stars.
Other factors that can contribute are a low stellar mass
or a low metallicity, although it is not clear whether these
factors can provide the low densities that are needed.
A related point is that some of the densities deduced around
Type Ic supernovae are surprisingly small.
For moderately efficient production of synchrotron emission
from SN 1998bw and SN 2001ap, the surrounding wind density
corresponds to $A_*\sim 0.01-0.05$ (Li \& Chevalier 1999;
Berger, Kulkarni, \& Chevalier 2002).
It is only if the fraction of energy density going into magnetic
fields ($\epsilon_B$) is small that the value of $A_*$ comes to
$\sim 1$.
At the same time, moderately high values of mass loss from
Wolf-Rayet stars are expected in order to obtain compatibility
with stellar evolution (e.g., Langer 1989).

We have discussed afterglow models in terms of interaction with
either a free wind or a constant density medium.
These cases are the simplest for modeling because the dynamical
situation can be treated in terms of a self-similar solution.
If both of these cases are relevant, there is also the expectation
that the GRB blast wave should in some cases be observed to
traverse the termination shock wave, which marks the transition
between the two types of media.
Wijers (2001) has described some of the basic changes that might
be expected for this transition, but they have not been clearly observed
in any afterglow, which is a point against the general scenario.
However, there is a need for more detailed hydrodynamic simulations
of this interaction, as well as calculations of
the expected emission properties.

\acknowledgments
RAC is grateful to Joe Cassinelli for correspondance on Wolf-Rayet
star winds and Brad Schaefer for discussions of absorption lines
in the spectra of GRB afterglows.
Support for this work was provided in part by NASA and the NSF,
and the Swedish Research Council.

\vfill\eject

\vspace*{1 cm}
\begin{center}
{\bf Table 1}

{\bf Free Wind Models for Afterglows}
\end{center}
\bigskip

\begin{tabular}{ccl}
\hline
GRB &  $A_*$   &  Reference  \\

\hline
 970508  & 0.3,0.39  &  CL00,PK02 \\
 991208  & 0.4,0.65  &  Li \& Chevalier 2001, PK02 \\
 991216  & $\sim 1$  & PK01  \\
 000301C  & 0.45  & Li \& Chevalier 2001  \\
 000418  & 0.69  & PK02  \\
 011121  &  0.02 & Price et al. 2002c  \\
 020405  & $\la 0.07$  &  this paper \\
 021004  & 0.6  &  Li \& Chevalier 2003 \\
 021211  & $0.0005$,$\sim 0.015$  & Kumar \& Panaitescu 2003; this paper  \\
\hline
\end{tabular}

\vfill\eject

\vspace*{1 cm}
\begin{center}
{\bf Table 2}

{\bf Afterglow properties}
\end{center}
\bigskip

\begin{tabular}{ccccccccc}
\hline
GRB &  $z$   &  $n$  &  $E_{53,iso}$  & $t_i/(1+z)$ & $p_0v_8^{-2}$ &
$R_t \dot M_{-5}^{-1/2}v_8^{1/2}$ & $\dot M_{-5}v_8^{-1}$ & $R_i/R_f$ \\
   &       &  cm$^{-3}$    &   (ergs)   &   days &  &  ($10^{18}$ cm) &
  $<$  &  $<$ \\
(1)  & (2)  & (3)  & (4)  & (5)  & (6)  & (7)  & (8) & (9)  \\
\hline
990123 & 1.60  &  0.0019 & 2.2 & 0.07 & 4.6(4) &28 & 0.0070 & 0.54\\
990510 & 1.62  & 0.29 & 0.96 & 0.08 & 7.0(6) & 2.3  &  0.060 & 0.64 \\
000301c & 2.03  & 27 &  0.11 & 0.17 & 6.5(8) & 0.24 & 0.29 & 0.55 \\
000926 & 2.07  & 22 & 0.32 & 0.26  & 5.3(8)  &  0.26 & 0.55 & 0.86 \\
\hline
\end{tabular}

\clearpage

\begin{figure}  
    \centering
    \leavevmode
    \psfig{file=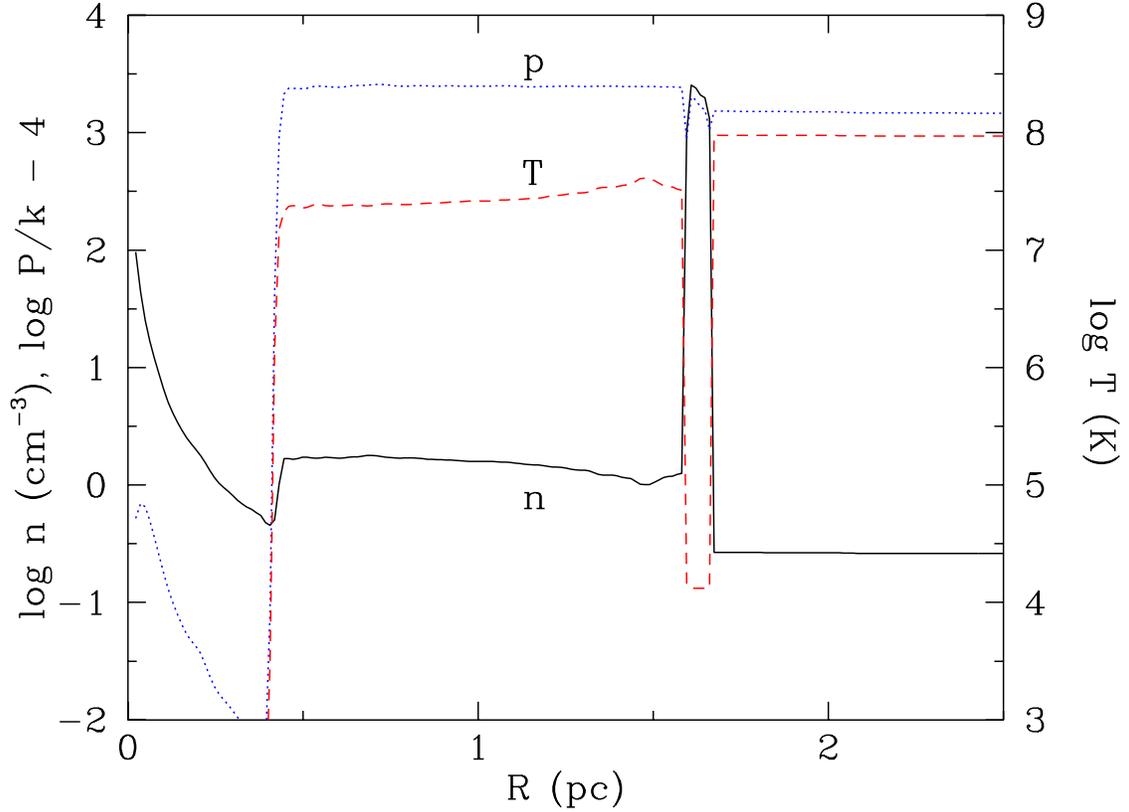,width=0.65\textwidth,angle=-90}
\caption{Wind bubble structure at the end of the
Wolf-Rayet stage for the case of an ISM pressure and density typical of the
hot, low density phase of a starburst galaxy, with $P/k = 2\times10^7 ~\rm K
~cm^{-3}$ and a density of $0.2 ~$ cm$^{-3}$. The solid line gives the
number density, the dashed line the temperature and the dotted line
the pressure.  The wind termination shock is at 0.4 pc
and the red supergiant shell at 1.7 pc. The region outside the red
supergiant shell is the remains of the bubble from the main sequence
phase. }
\label{fig1}
\end{figure}

\end{document}